# CONSERVED CURRENTS OF THE MAXWELL EQUATIONS
# WITH ELECTRIC AND MAGNETIC SOURCES

A. Gersten

Department of Physics, Ben-Gurion University, Beer-Sheva, Israel
e-mail: gersten@bgumail.bgu.ac.il

## ABSTRACT

New Lagrangians, depending on the field strengths and the electric and magnetic sources are found, which lead to the Maxwell equations. One new feature is that the equations of motion are obtained by varying the Lagrangian with respect to both the field strengths and the sources. In this way, conserved currents can be found for the field strengths and the electric or magnetic sources. Furthermore, using the equations of motion, the electric or magnetic sources can be eliminated, leading to conserved currents for the field strengths only (in the presence of electric and magnetic sources). Another new feature is the construction of a Lagrangian invariant under the duality transformation for both field strengths and electric and magnetic sources. The conserved current, after the elimination of electric and magnetic sources, depends on the field strengths only. The conserved quantity is related to the total helicity of the electromagnetic field.

1. <u>INTRODUCTION</u>

It is well known that the Maxwell equations were the source of inspiration for many important developments in physics. The symmetries of these equations led to dramatic discoveries. Yet it is still remarkable that new symmetries and conservation laws of the Maxwell equations are continuously being discovered [1-11]. Fushchich and Nikitin [1] have found and collected the most impressive amount of such symmetries and conservation laws. According to them, there is still a hope for new results "since Maxwell's equations have a hidden (non-geometrical) symmetry...".

More than 20 years ago, Lipkin [4] found unexpected conserved currents, which led Kibble and Fairlie [5] to develop a method generating an infinite number of conserved currents. Anderson and Arthurs [6] have derived a Lagrangian for the Maxwell equations depending on the field strengths and not the potentials. A similar Lagrangian was derived by Rosen [7]. The deficiency of this formalism is that this Lagrangian is the time component of a vector. Recently this formalism was improved by Sudbery [8] who generalized the previous Lagrangian to a vector, from which he deduced the conserved currents of Lipkin. His Lagrangian for the free Maxwell field was invariant under the duality transformation. The conserved quantities appear to be the symmetric energy momentum tensor.

In discussing the symmetries of the Maxwell equations, we should mention the attempts to present these equations as a first-quantized wave equation [1-3, 9-11]. These presentations also have the property of depending on the field strengths only.

In our work, new Lagrangians depending on the field strengths and electric and magnetic currents are derived. The new feature is that the equations of motion are obtained by varying the Lagrangians with respect to both field strengths and the electric (and optionally the magnetic) sources. In this way, the Lagrangians, although depending on the electric and magnetic currents, are not explicitly dependent on the co-ordinates; thus conserved currents can be found which include the fields strengths as well as the electric or magnetic currents. Furthermore, using the equations of motion, the electric or magnetic currents can be eliminated from the conserved currents. We thus obtain conserved currents for the field strengths only, which are valid even if electric or magnetic sources are present.

In our work we emphasize primarily the significance of the new method of obtaining conserved currents in the presence of electric and magnetic sources. The result is quite unexpected. Therefore we shall not yet concentrate on the significance of the new Lagrangians, the physical meaning of the conserved currents or quantization problems, leaving it for future consideration.

We also find in our work a Lagrangian invariant under the duality transformation of both field strengths and electric and magnetic sources (Section 3). The conserved quantity is related to the total helicity of the fields.

## 2. - THE FIRST SET OF NEW LAGRANGIANS

Throughout the paper we shall use the four-vector notation of Ref. [12]. The dual of an antisymmetric tensor $A_{\mu\nu}$ will be defined as $A_{\mu\nu}^D = \frac{1}{2}\varepsilon_{\mu\nu\sigma\lambda}A_{\sigma\lambda}$ and $\partial_\mu \equiv \partial/\partial x_\mu$, where $\varepsilon_{\mu\nu\sigma\lambda}$ is the totally antisymmetric Levi-Civita tensor (density). Note that $(A^D)^D = A$. The fields and currents will be functions of the four co-ordinates, and c stands for the velocity of light. The Maxwell equations are given through the antisymmetric electromagnetic field tensor $F_{\mu\nu}$:

$$\partial_\nu F_{\mu\nu} = -(4\pi/c)j_\nu^e \qquad (2.1a)$$

$$\partial_\mu F_{\mu\nu}^D = 0 \qquad (2.1b)$$

where $j_\nu^e$ are the electric currents ($j_4^e = ic\rho^e$, where $\rho^e$ is the electric charge density). The explicit form of $F_{\mu\nu}$ and $F_{\mu\nu}^D$ is given in Appendix A. From Eqs. (2.1), one can easily derive (see Appendix A) the following relation:

$$F_{\mu\nu} = -\tfrac{4\pi}{c}(\partial_\mu j_\nu^e - \partial_\nu j_\mu^e) \qquad (2.2)$$

i.e., the solutions of Eqs. (2.1) satisfy Eq. (2.2), but not always vice versa.

If magnetic sources are included, the Maxwell equations take the form [13]:

$$\partial_\mu F_{\mu\nu} = -(4\pi/c) j_\nu^e \qquad (2.3a)$$

$$\partial_\mu F_{\mu\nu}^D = -i(4\pi/c) j_\nu^m \qquad (2.3b)$$

where $j_\nu^m$ are the magnetic currents ($j_4^m = ic\rho^m$; $\rho^m$ is the magnetic charge distribution). From Eqs. (2.3), one can derive (see Appendix A) the following relation:

$$F_{\mu\nu} = -(4\pi/c)[\partial_\mu j_\nu^e - \partial_\nu j_\mu^e + i(\partial_\mu j_\nu^m - \partial_\nu j_\mu^m)^D] , \qquad (2.4)$$

and by taking the dual of this equation, we obtain the following consistent equation (with respect to $F \leftrightarrow F^D$, $j^e \leftrightarrow ij^m$):

$$F_{\mu\nu} = -(4\pi/c)[(\partial_\mu j_\nu^e - \partial_\nu j_\mu^e)^D + i(\partial_\mu j_\nu^m - \partial_\nu j_\mu^m)] , \qquad (2.5)$$

Let us construct the following Lagrangians, from which Eqs. (2.2) and (2.4) can be derived:

$$\mathcal{L}_I = -\tfrac{1}{2}(\partial_\mu F_{\lambda\sigma})(\partial_\mu F_{\lambda\sigma}) + \tfrac{4\pi}{c} F_{\mu\nu}(\partial_\mu j_\nu^e - \partial_\nu j_\mu^e) - (\tfrac{4\pi}{c})^2 j_\mu^e j_\mu^e , \qquad (2.6)$$

$$\mathcal{L}_{II} = \mathcal{L}_I + i\tfrac{4\pi}{c} F_{\mu\nu}(\partial_\mu j_\nu^m - \partial_\nu j_\mu^m)^D + (\tfrac{4\pi}{c})^2 j_\mu^m j_\mu^m , \qquad (2.7)$$

By varying $\mathcal{L}_I$ with respect to $F_{\mu\nu}$, Eq. (2.2) is obtained, whereas Eq. (2.1a) is derived by varying with respect to $j_\mu^e$. If we vary $\mathcal{L}_{II}$ with respect to $F_{\mu\nu}$, we obtain Eq. (2.4); Eq. (2.3a) is obtained when varying with respect to $j_\mu^e$, and Eq. (2.3b) when varying with respect to $j_\mu^m$.

The peculiarity of the Lagrangians $L_I$ and $L_{II}$ is that they do not depend explicitly on the co-ordinates; therefore we can obtain from them conserved currents (using Noether theorems). We may note that the solutions of the Maxwell equations (2.1) are solutions of the equations generated from $L_I$, but not always vice versa, but $L_{II}$ reproduces the Maxwell equations completely. Although $L_I$ may lead to solutions different from the Maxwell equations, still the conserved currents derived from $L_I$ are also conserved for the solutions of the Maxwell equations. This is because the solutions of the equations (2.1) are solutions of equations corresponding to $L_I$.

Let us construct for $L_I$ the following tensor:

$$T^I_{\mu\nu} = L_I \delta_{\mu\nu} - \frac{\partial L_I}{\partial(\partial_\nu F_{\sigma\lambda})} \partial_\mu F_{\sigma\lambda} - \frac{\partial L_I}{\partial(\partial_\nu j^e_\sigma)} \partial_\mu j^e_\sigma$$
$$= L_I \delta_{\mu\nu} + (\partial_\nu F_{\sigma\lambda})(\partial_\mu F_{\sigma\lambda}) - \frac{8\pi}{c} F_{\nu\sigma} \partial_\mu j^e_\sigma \quad (2.8)$$

for which, from Noether theorems,

$$\partial_\nu T^I_{\mu\nu} = 0 \ . \quad (2.9)$$

We can furthermore substitute into Eq. (2.8) the equations of motion (2.1a) and (2.2) and obtain:

$$T^I_{\mu\nu} = \delta_{\mu\nu}[-\tfrac{1}{2}(\partial_\eta F_{\sigma\lambda})(\partial_\eta F_{\sigma\lambda}) - F_{\eta\sigma}\partial_\lambda\partial_\lambda F_{\eta\sigma} - (\partial_\eta F_{\eta\lambda})(\partial_\sigma F_{\sigma\lambda})]$$
$$+ (\partial_\nu F_{\sigma\lambda})(\partial_\mu F_{\sigma\lambda}) + 2F_{\nu\sigma}\partial_\mu\partial_\lambda F_{\lambda\sigma} \quad (2.10)$$

In a similar way, we can construct from $L_{II}$ conserved currents from the tensor:

$$T^{II}_{\mu\nu} = L_{II}\delta_{\mu\nu} - \frac{\partial L_{II}}{\partial(\partial_\nu F_{\sigma\lambda})}\partial_\mu F_{\sigma\lambda} - \frac{\partial L_{II}}{\partial(\partial_\nu j^e_\sigma)}\partial_\mu j^e_\sigma - \frac{\partial L_{II}}{\partial(\partial_\nu j^m_\sigma)}\partial_\mu j^m_\sigma$$
$$= L_{II}\delta_{\mu\nu} + (\partial_\nu F_{\sigma\lambda})(\partial_\mu F_{\sigma\lambda}) - \frac{8\pi}{c}(F_{\nu\sigma}\partial_\mu j^e_\sigma + iF^D_{\nu\sigma}\partial_\mu j^m_\sigma) \quad (2.11)$$

Substituting in the above expression Eqs. (2.3) and (2.4), we obtain:

$$T^{II}_{\mu\nu} = \delta_{\mu\nu}[-\tfrac{1}{2}(\partial_\eta F_{\sigma\lambda})(\partial_\eta F_{\sigma\lambda}) - F_{\eta\sigma}\partial_\lambda\partial_\lambda F_{\eta\sigma} - (\partial_\eta F_{\eta\lambda})(\partial_\sigma F_{\sigma\lambda}) - (\partial_\eta F^D_{\eta\lambda})(\partial_\sigma F^D_{\sigma\lambda})]$$
$$+ (\partial_\nu F_{\sigma\lambda})(\partial_\mu F_{\sigma\lambda}) + 2F_{\nu\sigma}\partial_\mu\partial_\lambda F_{\lambda\sigma} + 2F^D_{\nu\sigma}\partial_\mu\partial_\lambda F^D_{\lambda\sigma} \quad (2.12)$$

## 3. - THE SELF-DUAL AND ANTI-SELF-DUAL CASES

Let us introduce the four-vector:

$$\Psi_a = H_a - iE_a, \quad a = 1,2,3; \qquad \Psi_4 = 0, \tag{3.1}$$

where $H_a$ and $E_a$ are the magnetic and electric field strengths respectively. Let us denote

$$q_\mu = \tfrac{4\pi}{c}(j_\mu^e + ij_\mu^m), \qquad d_\mu = \tfrac{4\pi}{c}(j_\mu^e - ij_\mu^m), \tag{3.2}$$

$$k_\mu = q_\mu^* + 2\partial_4 \Psi_\mu^*, \tag{3.3}$$

where $q_\mu^*$ and $\Psi_\mu^*$ are the complex conjugates of $q_\mu$ and $\Psi_\mu$ respectively. In Appendix B it is shown [Eqs. (B.14)] that the following equations are satisfied:

$$i(R_\mu^+)_{va}\partial_\mu \Psi_a = q_v, \tag{3.4}$$

$$i(R_\mu^+)_{va}\partial_\mu \Psi_a^* = k_v, \tag{3.5}$$

where the matrices $R_\mu^+$ are the Hermitian conjugates of $R_\mu$ given in Eq. (B.4). Each one of the above equations is equivalent to the Maxwell equations (2.3). One can check that the following relation is satisfied:

$$(-iR_\mu^+\partial_\mu)_{\alpha\beta}(iR_v\partial_v)_{\beta\gamma} = \delta_{\alpha\gamma}\partial_\mu\partial_\mu. \tag{3.6}$$

Multiplying Eqs. (3.4) and (3.5) by $-iR_\mu\partial_\mu$, we obtain

$$\Psi_a = -i(R_\mu)_{av}\partial_\mu q_v, \tag{3.7}$$

$$\Psi_a^+ = -i(R_\mu)_{av}\partial_\mu k_v. \tag{3.8}$$

One can construct the Lagrangian

$$L^{SD} = -\tfrac{1}{2}(\partial_\mu\Psi_a)(\partial_\mu\Psi_a) + i\Psi_a(R_\mu)_{av}\partial_\mu q_v + \tfrac{1}{2}q_v q_v, \tag{3.9}$$

which, when varied with respect to $\Psi_a$, gives Eq. (3.7), and when varied with respect to $q_v$, gives Eq. (3.4), after employing the relation

$$(R_\mu^+)_{av} = -(R_\mu)_{va}. \tag{3.10}$$

In a similar way, Eqs. (3.8) and (3.5) can be obtained from the Lagrangian

$$L^{ASD} = -\tfrac{1}{2}(\partial_\mu \Psi_a^*)(\partial_\mu \Psi_a^*) + i\Psi_a^*(R_\mu)_{av}\partial_\mu k_v + \tfrac{1}{2}k_v k_v \ , \qquad (3.11)$$

with conservation laws similar to the previous case. But the Lagrangian

$$L = -(\partial_\mu \Psi_a^*)(\partial_\mu \Psi_a) + i\Psi_a^*(R_\mu)_{av}\partial_\mu q_v + i\Psi_a(R_\mu)_{av}\partial_\mu k_v - k_v q_v \ , \qquad (3.12)$$

has an additional symmetry. Varied with respect to $\Psi_a$, $\Psi_a^*$, $q_v$, and $k_v$ it leads to Eqs. (3.7), (3.8), (3.4) and (3.5). It is invariant with respect to the following simultaneous duality transformations of the field strengths and sources:

$$\Psi_a \to \Psi_a e^{i\alpha}, \qquad \Psi_a^* \to \Psi_a^* e^{-i\alpha}, \qquad q_v \to q_v e^{i\alpha}, \qquad k_v \to k_v e^{-i\alpha}. \qquad (3.13)$$

Using Noether's theorem the following conserved current $J_\mu$ is obtained:

$$J_\mu = \tfrac{i}{2}(\Psi_a^* \partial_\mu \Psi_a - \Psi_a \partial_\mu \Psi_a^*) - \tfrac{1}{2}\Psi_a^*(R_\mu)_{av} q_v + \tfrac{1}{2}\Psi_a(R_\mu)_{av} k_v \ . \qquad (3\text{-}14)$$

In terms of the field strengths and electric and magnetic sources, it can be written as

$$\begin{aligned}
J_\mu &= H_a \partial_\mu E_a - E_a \partial_\mu H_a - 4\pi[\rho^m \vec{E} - \rho^e \vec{H} - (\vec{E} \times \vec{j}^e - \vec{H} \times \vec{j}^m)/c]_\mu \\
&= H_a \partial_\mu E_a - E_a \partial_\mu H_a - [(\nabla \cdot \vec{H})\vec{E} - (\nabla \cdot \vec{E})\vec{H} - \vec{E} \times (\nabla \times \vec{H} - \tfrac{1}{c}\tfrac{\partial \vec{E}}{\partial t}) - \vec{H} \times (\nabla \times \vec{E} + \tfrac{1}{c}\tfrac{\partial \vec{H}}{\partial t})]_\mu \ ; \\
\mu &= 1,2,3,
\end{aligned}$$

$$(3.15)$$

and the conserved density is:

$$\begin{aligned}
J_0 = -iJ_4 &= \vec{E} \cdot (\tfrac{1}{c}\tfrac{\partial \vec{H}}{\partial t} + \tfrac{4\pi}{c}\vec{j}^m) - \vec{H} \cdot (\tfrac{1}{c}\tfrac{\partial \vec{E}}{\partial t} + \tfrac{4\pi}{c}\vec{j}^e) \\
&= -\vec{E} \cdot (\nabla \times \vec{E}) - \vec{H} \cdot (\nabla \times \vec{H}).
\end{aligned} \qquad (3.16)$$

In the absence of sources, this result coincides with a conserved quantity found by Lipkin [4] [see also Refs. [6], [8] and [10]] for the free electromagnetic field. Calkin [14] has shown that this conserved quantity is proportional to the difference in the number of right and left circularly polarized photons. In other words, it can be regarded as proportional to the total helicity of the field. We found that it is conserved in the presence of the electric and magnetic sources.

It is interesting to note that fields with the property $\vec{E} \cdot (\nabla \times \vec{E}) = 0$ have been under investigation since the middle of the last century [15]. They are related to the so-called screw fields [15], [10] or Beltrami [15] vector fields, which have the property $\vec{E} \times (\nabla \times \vec{E}) = 0$.

## 4. - SUMMARY AND CONCLUSIONS

In our work we have succeeded in constructing Lagrangians which reproduce the Maxwell equations when varied with respect to both field strengths and sources. This leads to the finding of a new type of conservation laws in which the electric and magnetic sources are included. For the Lagrangians developed in this work it was possible, furthermore, to eliminate the sources from the conserved currents by using equations of motion [Eqs. (2.10) and (2.12)].

To the best of our knowledge, we have also been successful for the first time in constructing a Lagrangian [Eq. (3.12)] invariant under the duality transformation for both field strengths and electric and magnetic sources. The conserved current [Eqs. (3.14) and (3.16)1 is related to the conserved current found previously for the free electromagnetic field by Lipkin; but in our case, it is also conserved in the presence of electric and magnetic sources.

The new method developed by us for constructing conserved currents can be interpreted as a successful attempt to present the Maxwell equations with electric and magnetic sources as a closed system, irrespective of the exact nature of the sources.

## APPENDIX A

The Explicit Form of $F_{\mu\nu}$ and the Derivation of Eqs. (2.2) and (2.4)

Using the notation of Ref. [12], the electromagnetic field antisymmetric tensor $F_{\mu\nu}$ is given via the electric and magnetic field strengths $\vec{E}$ and $\vec{H}$ respectively by:

$$(F) = \begin{pmatrix} 0 & H_z & -H_y & -iE_x \\ -H_z & 0 & H_x & -iE_y \\ H_y & -H_x & 0 & -iE_z \\ iE_x & iE_y & iE_z & 0 \end{pmatrix}. \quad (A.1)$$

The dual tensor $F_{\mu\nu}^D = \frac{1}{2}\varepsilon_{\mu\nu\lambda\sigma}F_{\lambda\sigma}$ is given by

$$(F^D) = \begin{pmatrix} 0 & -iE_z & iE_y & H_x \\ iE_z & 0 & -iE_x & H_y \\ -iE_y & iE_x & 0 & H_z \\ -H_x & -H_y & -H_z & 0 \end{pmatrix}. \quad (A.2)$$

Equation (2.1b) can be presented also in the following well-known way:

$$\partial_\mu F_{\lambda\sigma} + \partial_\sigma F_{\mu\lambda} + \partial_\lambda F_{\sigma\mu} = 0. \quad (A.3)$$

Taking the derivative of Eq. (A.3), we obtain

$$\partial_\mu \partial_\mu F_{\lambda\sigma} = -\partial_\mu(\partial_\sigma F_{\mu\lambda} + \partial_\lambda F_{\sigma\mu}) = -\partial_\sigma \partial_\mu F_{\mu\lambda} + \partial_\lambda \partial_\mu F_{\mu\sigma}. \quad (A.4)$$

Now we substitute on the right-hand side of Eq. (A.4) the Maxwell equation (2.1a) and obtain Eq. (2.2). One can check that Eqs. (2.3) are equivalent to

$$\partial_\mu F_{\lambda\sigma}^D + \partial_\sigma F_{\mu\lambda}^D + \partial_\lambda F_{\sigma\mu}^D = -\frac{4\pi}{c}\varepsilon_{\lambda\sigma\mu\nu}j_\nu^e, \quad (A.5a)$$

$$\partial_\mu F_{\lambda\sigma} + \partial_\sigma F_{\mu\lambda} + \partial_\lambda F_{\sigma\mu} = -i\frac{4\pi}{c}\varepsilon_{\lambda\sigma\mu\nu}j_\nu^m. \quad (A.5b)$$

Applying the $\partial_\mu$ derivative from the left to Eq. (A.5a), we obtain

$$\partial_\mu \partial_\mu F_{\lambda\sigma}^D = \partial_\lambda \partial_\mu F_{\mu\sigma}^D - \partial_\sigma \partial_\mu F_{\mu\lambda}^D - \frac{4\pi}{c}\varepsilon_{\lambda\sigma\mu\nu}\partial_\mu j_\nu^e. \quad (A.6)$$

Using

$$\varepsilon_{\lambda\sigma\mu\nu}\partial_\mu j_\nu^e = \frac{1}{2}\varepsilon_{\lambda\sigma\mu\nu}(\partial_\mu j_\nu^e - \partial_\nu j_\mu^e) = (\partial_\lambda j_\sigma^e - \partial_\sigma j_\lambda^e)^D, \quad (A.7)$$

and substituting Eq. (2.3b) into the right-hand side of Eq. (A.6), we prove the relation (2.5). The relation (2.4) can be proved in a similar way by applying to Eq. (A.5b) the $\partial_\mu$ derivative from the left.

**APPENDIX B**

The Self-Dual and Anti-Self-Dual Equations

From the tensors (A.1) and (A.2) we can form the self-dual combination

$$F^{SD} = F + F^D ,\qquad(B.1)$$

and the anti~self-dual combination

$$F^{ASD} = F - F^D .\qquad(B.2)$$

Using the definition (3.1), the $F^{SD}$ and $F^{ASD}$ are given by

$$(F^{SD}) = \begin{pmatrix} 0 & \Psi_z & -\Psi_y & \Psi_x \\ -\Psi_z & 0 & \Psi_x & \Psi_y \\ \Psi_y & -\Psi_x & 0 & \Psi_z \\ -\Psi_x & -\Psi_y & -\Psi_z & 0 \end{pmatrix}, \qquad (F^{ASD}) = \begin{pmatrix} 0 & \Psi_z^* & -\Psi_y^* & -\Psi_x^* \\ -\Psi_z^* & 0 & \Psi_x^* & -\Psi_y^* \\ \Psi_y^* & -\Psi_x^* & 0 & -\Psi_z^* \\ \Psi_x^* & \Psi_y^* & \Psi_z^* & 0 \end{pmatrix}, \qquad(B.3)$$

Let us define the two sets of matrices

$$S_1 = \begin{pmatrix} & & i & \\ & & & i \\ & -i & & \\ -i & & & \end{pmatrix},\ S_2 = \begin{pmatrix} & & -i & \\ & & & i \\ i & & & \\ & -i & & \end{pmatrix},\ S_3 = \begin{pmatrix} & i & & \\ -i & & & \\ & & & i \\ & & -i & \end{pmatrix},\ S_4 = \begin{pmatrix} i & & & \\ & i & & \\ & & i & \\ & & & i \end{pmatrix},$$

$$R_1 = \begin{pmatrix} & & & -i \\ & & i & \\ & -i & & \\ i & & & \end{pmatrix},\ R_2 = \begin{pmatrix} & & -i & \\ & & & -i \\ i & & & \\ & i & & \end{pmatrix},\ R_3 = \begin{pmatrix} & i & & \\ -i & & & \\ & & & -i \\ & & i & \end{pmatrix},\ R_4 = \begin{pmatrix} -i & & & \\ & -i & & \\ & & -i & \\ & & & -i \end{pmatrix},$$

(B.4)

then

$$F_{\mu\nu}^{SD} = -i(S_a)_{\mu\nu}\Psi_a ,\qquad a=1,2,3,\qquad(B.5)$$

$$F_{\mu\nu}^{ASD} = -i(R_a)_{\mu\nu}\Psi_a^* ,\qquad a=1,2,3.\qquad(B.6)$$

Furthermore, we assume [9]

$$\Psi_4 = \Psi_4^* = 0 ,\qquad(B.7)$$

and note that

$$(R_\mu)_{va} = (S_a^+)_{\mu v} ,  \qquad (B.8)$$

and equivalently

$$(S_\mu)_{va} = (R_a^+)_{\mu v} .  \qquad (B.9)$$

Then, from Eqs. (2.3), (3.2) and (B.5)-(B.9), we have

$$\partial_\mu F_{\mu v}^{SD} = -i(S_a)_{\mu v}\partial_\mu \Psi_a = -i(R_\mu^+)_{va}\partial_\mu \Psi_a = -q_v ,  \qquad (B.10)$$

$$\partial_\mu F_{\mu v}^{ASD} = -i(R_a)_{\mu v}\partial_\mu \Psi_a^* = -i(S_\mu^+)_{va}\partial_\mu \Psi_a^* = -d_v .  \qquad (B.11)$$

A few remarks are needed here. Equations (B.10) and (B.11) are consistent with each other, but as $\Psi_4 = \Psi_4^* = 0$, there exists an ambiguity in defining $(R_\mu)_{v4}$ and $(S_\mu)_{v4}$. This is the reason why Eq. (B.11) is not explicitly the complex conjugate of (B.10), but one can check that they are consistent. Equation (B.11) can explicitly be made the complex conjugate of Eq. (B.10), if it is replaced by:

$$i(R_\mu)_{va}\partial_\mu \Psi_a^* = q_v^* .  \qquad (B.12)$$

We can also rewrite it as follows:

$$q_v^* = i(R_\mu^+)_{va}\partial_\mu \Psi_a^* - 2i(R_4)_{va}\partial_4 \Psi_a^* ,$$

or, using Eq. (3.3):

$$i(R_\mu^+)_{va}\partial_\mu \Psi_a^* = q_v^* + 2\partial_4 \Psi_v^* \equiv k_v .  \qquad (B.13)$$

Thus finally the two complex conjugate equations can be written as:

$$i(R_\mu^+)_{va}\partial_\mu \Psi_a = q_v ,  \qquad (B.14a)$$

$$i(R_\mu^+)_{va}\partial_\mu \Psi_a^* = k_v .  \qquad (B.14b)$$

Equations (B.10) and (B.12) can be derived from the following Lagrangian density:

$$L_0 = \tfrac{i}{2}[\Psi_v^*(R_\mu^+)_{va}\partial_\mu \Psi_a + \Psi_v(R_\mu)_{va}\partial_\mu \Psi_a^*] - \Psi_v^* q_v - \Psi_v q_v^* .  \qquad (B.15)$$

For the free field equations ($q_\mu = q_\mu^* = 0$), translation invariance leads to the conserved tensor

$$\tau_{\mu\nu} = L_0 \delta_{\mu\nu} - \tfrac{i}{2}[\Psi_b^*(R_\nu^+)_{ba}\partial_\mu\Psi_a + \Psi_b(R_\nu)_{ba}\partial_\mu\Psi_a^*] \,, \tag{B.16}$$

which is related to the one found in Ref. [10]. The duality transformation

$$\Psi_a \to \Psi_a e^{i\alpha}, \qquad \Psi_a^* \to \Psi_a^* e^{-i\alpha},$$

leads to the conserved energy current

$$J_\mu = \tfrac{i}{2}[\Psi_b^*(R_\mu^+)_{ba}\Psi_a - \Psi_b(R_\mu)_{ba}\Psi_a^*] \,, \tag{B.17}$$

which is related to the one found in Ref. [8].